\documentclass[%
 reprint, %
superscriptaddress,
 amsmath,amssymb,
 aps, physrev,
 twocolumn,
]{revtex4-2}

\usepackage{graphicx}%
\usepackage{dcolumn}%
\usepackage{bm}%
\usepackage{hyperref}%
\usepackage{xcolor} %

\begin{document}

\preprint{APS/123-QED}

\title{\textbf{The impact of internal electric fields on accurate diamond magnetometry} 
}%

\author{Jack Faller}
\affiliation{Department of Physics, McGill University, 3600 Rue University, Montreal Quebec H3A 2T8, Canada}

\author{Daphne Freda Gavras}
\affiliation{Department of Physics, McGill University, 3600 Rue University, Montreal Quebec H3A 2T8, Canada}
\affiliation{Institute of Physics, \'Ecole Polytechnique Fédérale de Lausanne (EPFL), CH-1015 Lausanne, Switzerland}

\author{Andrew Lowther}
\affiliation{
 SBQuantum, 805 Rue Galt O, Sherbrooke, Quebec J1H 1Z1, Canada
}
\author{Romain Ruhlmann}
\affiliation{
 SBQuantum, 805 Rue Galt O, Sherbrooke, Quebec J1H 1Z1, Canada
}

\author{Lilian Childress}
\email{lilian.childress@mcgill.ca}
\affiliation{Department of Physics, McGill University, 3600 Rue University, Montreal Quebec H3A 2T8, Canada}

\date{\today}%

\begin{abstract}

Nitrogen-vacancy (NV) center spin ensembles are a promising platform for accurate vector magnetometry, as they offer measurements referenced to a spin Hamiltonian in an anisotropic crystal with native vector sensing capabilities. %
However,  charged dopants within the diamond produce random internal electric fields (RIEFs) that shift NV spin-transition frequencies, and their impact on the accuracy of NV magnetometry has not been investigated. Here, we characterize RIEFs' effect on NV spin-transition spectra and propose an analytical fitting function to improve accuracy in their presence. We build a Monte Carlo RIEF model that reproduces the asymmetric lineshapes observed in experimental optically detected magnetic resonance (ODMR) spectra, and use it as a testbed to quantify RIEF-induced errors by simulating ODMR spectra including RIEFs and comparing reconstructed magnetic fields to the known model input.  With standard analysis approaches, we observe systematic errors exceeding the g-factor uncertainty in fields below $\sim$ 5 mT, while the proposed fitting function reduces the error by nearly an order of magnitude. Our results demonstrate that RIEFs can play a significant role in the systematic error budget for NV magnetometry, and offer a path towards improved accuracy in NV quantum sensing. %
\end{abstract}

\maketitle

\section{Introduction}
 Over the last two decades, nitrogen-vacancy (NV) centers in diamond have become a leading system for precision DC magnetometry~\cite{barry_sensitivity_2020}. Owing to a long-lived ground-state $S = 1$ spin, which remains highly coherent even at room temperature, and spin-dependent optical properties, which permit high-contrast spin resonance detection, Zeeman-shifted NV spin transitions can be readily used to infer the magnetic field. Continuing advances in quantum control~\cite{bauch_ultralong_2018} and spin detection~\cite{hopper_spin_2018} have enabled near-DC NV magnetometry with sensitivity below pT/$\sqrt{\text{Hz}}$~\cite{barry_sensitive_2024, wang_spin-refrigerated_2024}. %
 
 While sensitivity is essential for research probing tiny magnetic signals such as those produced by neuron activation~\cite{barry_optical_2016}, in other magnetometry contexts including geomagnetic mapping~\cite{alken_international_2021}, navigation~\cite{canciani_airborne_2017}, resource exploration~\cite{nabighian_historical_2005} or space missions~\cite{bennett_precision_2021}, accuracy and stability become a primary concern. Currently, the most accurate scalar commercial magnetometers offer 50 pT absolute accuracy~\cite{noauthor_gem_nodate}, while the European Space Agency's earth-field-mapping SWARM mission has observed deployed accuracy near a nanotesla for full vector measurements~\cite{hulot_swarms_2015}, both using atomic magnetometers. With a spin-based operating mechanism similar to atomic sensors and the capability to intrinsically measure vector fields~\cite{maertz_vector_2010}, NV ensembles offer an intriguing alternative for high-accuracy magnetic sensing.

 Nevertheless, existing literature on NV accuracy is limited. NV magnetometer stability, which is a pre-requisite for accuracy, has been demonstrated below 1 pT over hours-long timescales~\cite{sekiguchi_diamond_2024, kainuma_compact_2024}. To date, there is only one reported experimental characterization of NV magnetometry in a nT-accuracy magnetic field environment, where a machine-learning calibration algorithm yielded $<10$~nT residuals over the full range of earth's magnetic field~\cite{halde_who_2025}, reaching an accuracy competitive with existing commercial portable vector magnetometers~\cite{noauthor_fgm3d_nodate}. Notably, an empirical calibration rather than first-principles analysis was required to obtain good accuracy.  The same team also observed asymmetries in spin resonance lineshapes that hindered identification of spin transition frequencies. 
These observations point to the existence of stable, well-defined spin transition frequencies in the NV, but indicate that current models fail to fully capture their behavior. Combined with the growing recognition of accuracy as a key figure of merit for NV magnetometry~\cite{lonard_limits_2026, leclerc_beyond_2026}, they motivate a more careful examination of sources of systematic error.

In this context, a recent insight into the NV spin environment deserves attention. In zero magnetic field, the nominally degenerate $m_s = \pm1$ NV spin sublevels are commonly observed to be split. For many years, the splitting was attributed to strain, but its behavior is more consistent with random internal electric fields (RIEFs) arising from charged dopants within the diamond ~\cite{mittiga_imaging_2018, yu_optically_2024, sorensen_nanophotonic_2025}. Such fields are necessarily present because the negatively charged NV obtains its extra electron by ionizing a nearby dopant, producing strong local electric fields that can exceed 10 MV/m in dense ensembles. The component of the electric field perpendicular to the NV axis ($E_\perp$) couples the $m_s = \pm1$ states, hybridizing them and splitting the hybridized states by $\mathcal{E}\propto E_\perp$. A sufficiently large axial magnetic field $B_z$ will suppress hybridization, and in the large-$B_z$ limit the electric-field-induced energy shift falls off as $\mathcal{E}/B_z$. Nevertheless, for vector magnetometry only moderate bias fields are typically employed to avoid spin-mixing that reduces signal contrast, and reduced or even zero bias field is desirable to study magnetically sensitive materials~\cite{zheng_zero-field_2019} or maximize stability~\cite{childress_bias-field-free_2025}. %
While it is well understood that RIEFs suppress magnetic sensitivity at zero field, their residual impact on accuracy at low to moderate bias magnetic field remains unexplored.

Here, we examine the role that RIEFs play in limiting the accuracy of finite-bias NV magnetometry. In particular, we seek to quantify the systematic error introduced by current data analysis approaches that neglect RIEFs. We build a numerical model and show that it is consistent with the asymmetric lineshapes observed in finite-bias optically detected magnetic resonance (ODMR) spectra. We then use this numerical model as a testbed, analyzing simulated data and comparing conventionally extracted magnetic fields with the model ground truth to characterize RIEF-induced systematic errors.
We find that these errors fall off more slowly with magnetic field than expected, $\sim 1/\sqrt{Bz}$; for a typical defect density (corresponding to commercially available samples) in few-mT bias fields, they exceed %
 the g-factor uncertainty, which was recently highlighted as an important source of systematic error~\cite{lonard_limits_2026}. %
We also introduce an alternative fitting function based on an approximate RIEF model that can reduce the RIEF-induced systematic error by an order of magnitude.
Our results indicate that RIEFs are a significant factor in the systematic error budget for NV magnetometry, and will need to be included in any attempt to reach absolute accuracy approaching empirical calibrations.

\section{Modeling ODMR with random internal electric fields}
\subsection{The probability distribution of internal electric fields}

\begin{figure}[hbt]
\includegraphics[width=\linewidth]{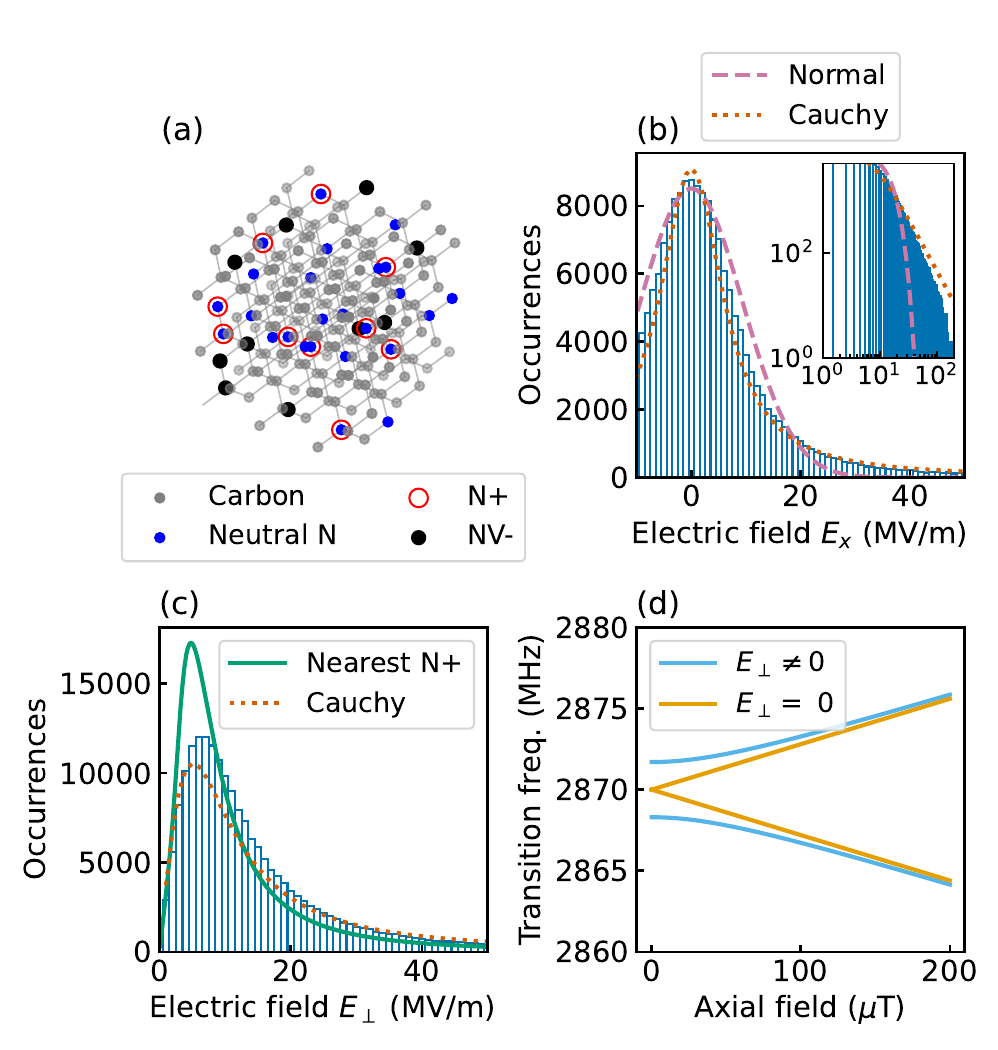}%
\caption{\label{fig1}(a) An illustration of the Monte Carlo NV$^-$-N$^+$ pair model (generated with code from Claude Sonnet 4.6). Randomly located NV$^-$ defects ionize the nearest nitrogen (also randomly distributed), producing a random total electric field at the location of the target NV$^-$. (b) A histogram of total electric field components $E_x$ observed for [NV$^-$] = 2.7~ppm, [N] = 8.5~ppm, along with fits to Cauchy and Normal distributions. Inset: Same data on a log-log axis illustrating a power-law tail. (c) Histogram of $E_\perp = \sqrt{E_x^2 + E_y^2}$ compared to the result from Cauchy-distributed $E_x$ and $E_y$ and to a model considering only the electric field from the nearest ionized N$^+$. (d) Illustration of the impact of a typical transverse electric field (10 MV/m) on the spin transition frequencies of the NV$^-$.}
\end{figure}

We use Monte Carlo (MC) simulation %
to investigate the impact of internal electric fields on inhomogeneously broadened spin spectra. Building on early MC models that surrounded each NV by a random distribution of positive and negative charges~\cite{mittiga_imaging_2018}, our approach follows Yu \textit{et al.}~\cite{yu_optically_2024} in adopting a NV$^-$-N$^+$ pair model~\cite{manson_nvn_2018}. We place N and NV$^-$ randomly within a diamond lattice centered on a fixed NV$^-$, with each NV$^-$ receiving its electron from the nearest neutral N, converting it to N$^+$ (see Fig.~\ref{fig1}(a)). We then calculate the total electric field at the central NV$^-$ from all the charged dopants (see Appendix~\ref{app:MC} for more details).

Figure~\ref{fig1}(b) shows the resulting distribution of internal electric fields for a moderately dense NV ensemble. The fields, on the order of kV/mm, are significantly larger than typical ambient electric fields present in a laboratory setting.  The distribution of each electric field component reveals a long Cauchy-like power-law tail, incompatible with the Gaussian distribution sometimes used as an approximation ~\cite{zhang_blueprint_2024}. Figure~\ref{fig1}(c) shows the distribution of electric fields perpendicular to the NV axis, which dominate the NV spin response due to a strongly anisotropic susceptibility~\cite{van_oort_electric-field-induced_1990}. This $E_\perp$ distribution is skewed to higher fields than predicted by a model that includes only the closest $N^+$ ion~\cite{yu_optically_2024}; it is somewhat better approximated by $\sqrt{E_x^2 + E_y^2}$ with $E_{x,y}$ drawn from identical Cauchy distributions. Since we were unable to find a simple analytic expression exhibiting close agreement with numerics, we continue to use the  MC simulation going forward.

The random distribution of internal fields impacts the NV spin spectrum primarily through  $E_\perp$, which partially hybridizes and further separates the $m_s = \pm1$ energy levels (see Fig.~\ref{fig1}(d)). While the energy shift is largest at zero external magnetic field, it persists to higher fields. Because $E_\perp$ can only increase the $m_s = \pm1$ separation, a distribution of $E_\perp$ values not only broadens but also shifts the line centers of the inhomogeneously broadened spin spectrum.

\subsection{Inhomogeneously broadened spin spectra}
The NV spin transition frequencies are governed by a Hamiltonian involving electronic ($\mathbf{S}$) and nuclear spin ($\mathbf{I}$) degrees of freedom:
    \begin{eqnarray}
        H/h &=& (\Delta + d_\parallel E_z) S_z^2 + \gamma_{NV} \mathbf{B}\cdot\mathbf{S}   \label{eq:H}\\
         \nonumber&& + d_\perp E_x\left(S_x^2 - S_y^2\right) + d_\perp E_y \left(S_x S_y+S_y S_x\right)\\
         \nonumber&& + d'E_x\left(S_x S_z + S_z S_x\right) + d' E_y\left(S_y S_z + S_z S_y\right)\\
         \nonumber&& + P I_z^2 - \gamma_N \mathbf{B}\cdot\mathbf{I} + A_\parallel S_z I_z + A_\perp \left(S_x I_x + S_y I_y\right).
    \end{eqnarray}
Here, $\hat{z}$ is parallel to the NV symmetry axis, $\Delta \approx 2.87$~GHz~\cite{doherty_nitrogen-vacancy_2013} is the zero-field splitting, $\gamma_{NV}  \approx 28$~GHz/T~\cite{rondin_magnetometry_2014} is the NV gyromagnetic ratio, $\mathbf{B}$ is the magnetic field, $d_\parallel \approx 0.0035$~Hz/(V/m), $d_\perp \approx 0.17$~Hz/(V/m)~\cite{van_oort_electric-field-induced_1990}, and $d'$ (unknown, but believed to be of a similar order of magnitude as $d_\perp$~\cite{doherty_theory_2012}) are the electric field susceptibilities, $\mathbf{E}$ is the electric field, $P \approx - 4.945$~MHz~\cite{steiner_universal_2010} is the $^{14}$N nuclear quadrupole splitting, $\gamma_N \approx 3.0777 $ MHz/T~\cite{harris_nmr_2002} is the $^{14}$N nuclear gyromagnetic ratio, and $A_\parallel\approx -2.16 $~MHz and $A_\perp\approx -2.7$ MHz~\cite{felton_hyperfine_2009} are the hyperfine coupling constants. $S_{x,y,z}$ and $I_{x,y,z}$ are dimensionless spin-1 angular momentum operators for the electronic spin and $^{14}$N nuclear spin respectively. In what follows, we neglect the effect of the terms involving $d'$ because they are strongly suppressed by the large zero-field splitting (whereas $d_\perp$ terms are suppressed only by the Zeeman splitting)~\cite{van_oort_electric-field-induced_1990}.

We use the simulated distribution of internal electric fields to construct a distribution of spin Hamiltonians, each representing an NV in the ensemble. The magnetic and electric fields are transformed into the appropriate coordinate system for the NV orientation under consideration, and we account for magnetic noise by adding a small random magnetic field $\delta\mathbf{B}$. Each component of $\delta\mathbf{B}$ is drawn from a Cauchy distribution with half width at half maximum (HWHM) given by $1/(2\pi T_2^*\gamma_{NV})$, where $T_2^*$ is the magnetic dephasing time~\cite{abragam_principles_1961}.

We construct the spin spectrum from the Hamiltonian distribution as a histogram of the electron spin transition frequencies. We incorporate nuclear spin structure within the secular approximation, in which only the $A_\parallel I_z$ hyperfine term is retained. Effectively, the nuclear spin produces an additional magnetic field $\mathbf{B_{\rm eff}} = m_I A_\parallel/\gamma_{NV} \hat{z}$, where $m_I$ has equal probability of taking on the values $\{-1, 0, 1\}$.  This approximation neglects the nonsecular off-axis hyperfine terms, which give rise to the weakly allowed transitions visible in experiment (Figs.~\ref{fig2}(b)--\ref{fig2}(e)) and lead to small shifts in transition frequencies (see Appendix~\ref{app:hyperfine}). Including these terms requires the full $9 \times 9$ Hamiltonian, an approach that is significantly more computationally demanding and unnecessary for probing the effects of RIEFs.

The bare spectrum of spin transition frequencies represents the signal that would be observed in optically detected magnetic resonance (ODMR) in the limit of low microwave (MW) and optical power; alternatively, it could be revealed by Ramsey spectroscopy in the limit of hard pulses. To account for MW broadening in ODMR, we convolve the simulated distribution of transition frequencies with a Lorentzian lineshape of half-width equal to the Rabi frequency. This adjustment, which neglects the full spin-1 dynamics, is appropriate when the bias field is sufficiently large that the MW field drives a single transition at a time.

\subsection{Comparison to experimental spectra}

Before using the Monte Carlo model to understand
the role electric fields play in magnetometry, we verify that it is consistent with experiment. Continuous-wave (CW) ODMR data were taken using the SBQuantum SBQDM device~\cite{halde_who_2025}, which utilizes a tunable MW cavity and $\sim$100 mW 520 nm laser to drive and monitor spin transitions within an Element Six sample. The device includes magnets that impose a 3--4~mT bias field in approximately the $\langle5\,2\,1\rangle$ direction. Owing to imperfections in the MW cavity tuning, the Rabi frequency varied across the MW frequency range probed. A small drifting background, likely due to laser power fluctuations, was subtracted (see Appendix~\ref{app:fit}).

\begin{figure}[htb]
\includegraphics[width=\linewidth]{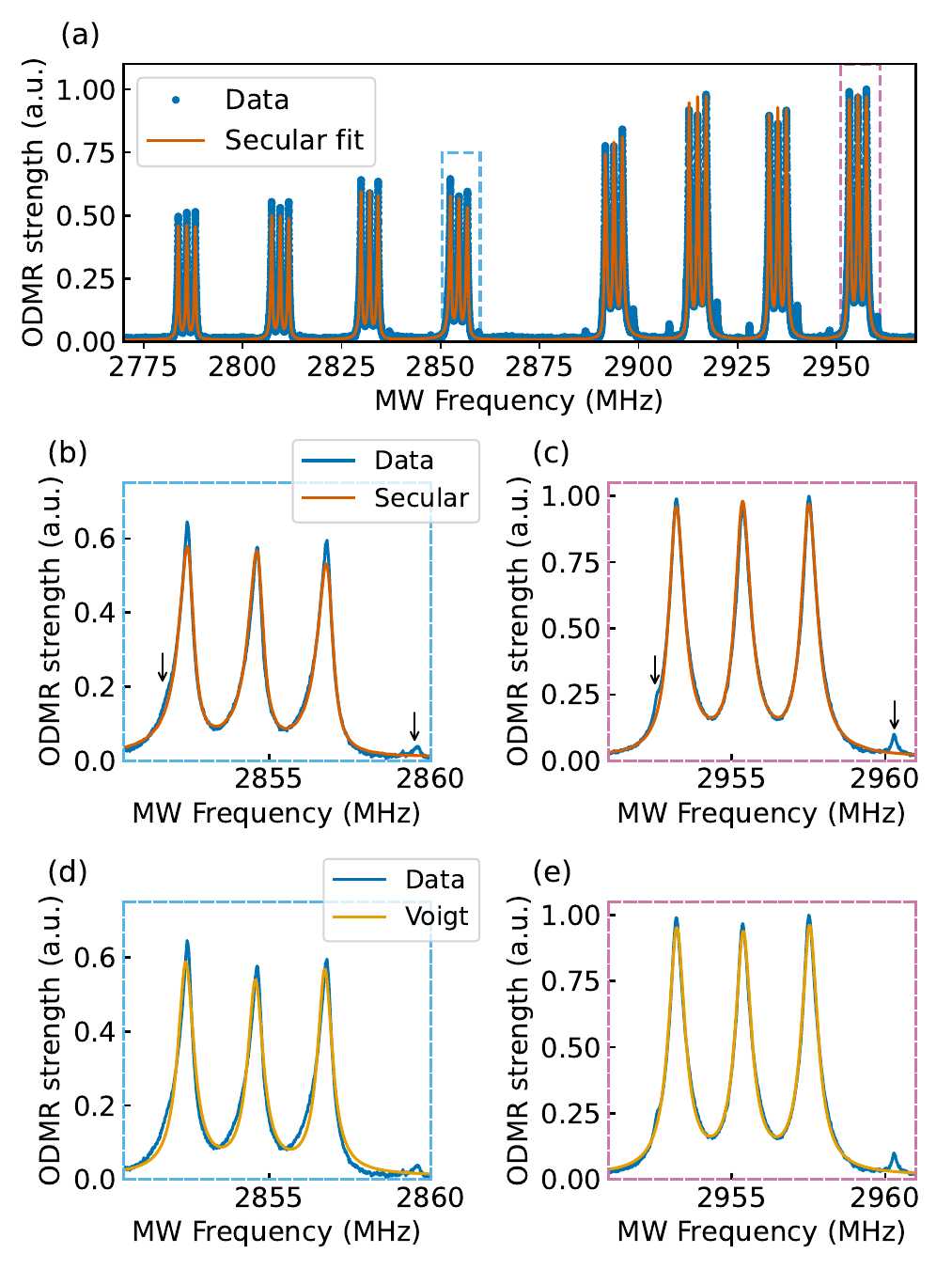}%
\caption{\label{fig2} (a) Optically detected magnetic resonance data of the Element Six sample, with background subtracted (blue). The orange line shows a fit to a numerically generated spectrum in the secular approximation, including internal electric fields from the nominal concentrations of NV$^-$ (2.7 ppm) and nitrogen (8.5 ppm) in the sample, see Appendix~\ref{app:fit} for details. (b,c) Details of the 4th (b) and 8th (c) triplets. Features closer to the central zero-field splitting exhibit more pronounced asymmetry, which is captured by the model. Weakly allowed transitions not included in the secular approximation are marked by black arrows. (d,e) Voigt fits to the same data. The fit fails to capture the asymmetry of the 4th triplet (d), while the more symmetric 8th triplet (e) is fit well.}
\end{figure}

We fit the data to the MC spectrum model in the secular approximation, including MW broadening with a different Rabi frequency for each triplet and a different contrast for each of the four NV orientations to account for unknown optical and MW polarization (see Appendix~\ref{app:fit} for details). The magnetic field, magnetic broadening $T_2^* \approx 1.24~\mu$s, and zero-field splitting were extracted from the fit. We fixed the NV$^-$ and N concentrations to the nominal specifications for the diamond sample, assuming that 60\% of the NV centers were in the negative charge state~\cite{thalassinos_robust_2025}. Figure~\ref{fig2}(a) shows the data and the fit. Notably, a detailed comparison of one of the inner triplets (Fig.~\ref{fig2}(b)) shows a distinct asymmetry in the data, which is well captured by the model; the outermost triplets (Fig.~\ref{fig2}(c)) show significantly less asymmetry, which also appears in the model. For comparison, a Voigt profile cannot replicate the asymmetry of the inner triplets (Fig.~\ref{fig2}(d),(e)). Some discrepancies remain, notably lambda systems in the $m_I = \pm 1$ transitions and nuclear-spin-nonconserving transitions (black arrows in Fig.~\ref{fig2}(b),(c)) that arise from nonsecular terms in the Hamiltonian. %

Importantly, the model predicts the asymmetric lineshapes observed in the data and reproduces the reduction in asymmetry going from inner to outer triplets. Such asymmetry has appeared in other published ODMR spectra~\cite{zheng_zero-field_2019, barry_sensitive_2024}, but it was not commented on and its source has not been clearly identified. The observed consistency with our model suggests that internal electric fields are playing an important role in ODMR, even in the presence of a bias field.

\section{Single-axis magnetometry} \label{sec:single}

To gain insight into the impact of internal electric fields on magnetometry, we first examine the implications of our model in the simplest scenario: a single NV orientation (with symmetry axis parallel to the magnetic field) and no hyperfine structure. Neglecting hyperfine structure isolates the effects coming from internal electric fields, and speeds computation.

We begin by focusing on the inhomogeneously broadened spin spectrum without additional microwave or optical broadening, as this reflects the inherent information present in the ensemble. Unless otherwise specified, we consider an NV ensemble similar to the experimental data in Fig.~\ref{fig2} with [N] = 8.5 ppm, [NV$^-$] = 2.7 ppm, and $T_2^* = 1.24~\mu$s.

\subsection{Lineshape asymmetry and peak shifts}

\begin{figure}[htb]
\includegraphics[width=\linewidth]{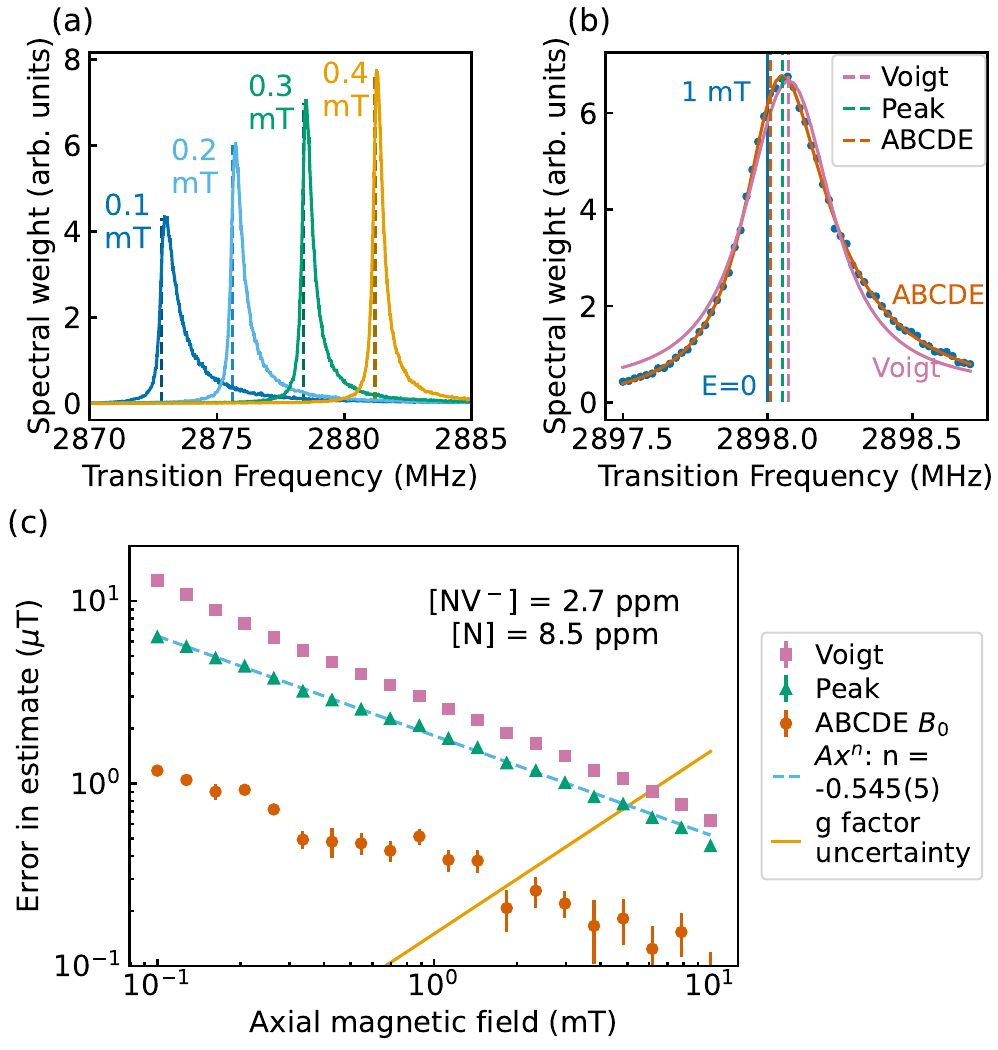}%
\caption{\label{fig3} (a) Evolution of lineshape and peak shift with axial magnetic field. Dashed lines illustrate the nominal spin transition frequency $f_0 = \Delta + \gamma_{NV} B_z$. (b) Comparison of fit functions at $B_z$ = 1 mT. Vertical lines indicate the nominal spin transition frequency $f_0$ (solid blue) as well as the center frequency $f_\text{Voigt}$ of a Voigt fit (dashed pink), $f_\text{ABCDE} = \Delta +\mathcal{B}_0$ from an ABCDE fit (dashed orange), and the maximum of the ABCDE fit $f_\text{peak}$ (dashed green). (c) Errors in the magnetic field estimated from $(f_i - \Delta)/\gamma_{NV}$ for $i = $ (Voigt, peak, ABCDE) compared to uncertainty arising from $g = 2.0028 \pm 0.0003$ (orange line). The dashed line shows a fit to a power law model $A B_z^ n$ with $n = -0.545\pm 0.005$. Error bars  are found by bootstrapping (see Appendix~\ref{app:bootstrapping}). %
In all panels, %
simulations use $[NV^-]$ = 2.7 ppm, $[N]$ = 8.5 ppm, and magnetic $T_2^* = 1.24~\mu$s.   %
}
\end{figure}

The first key feature introduced by internal electric fields is a pronounced asymmetry in the lineshape of inhomogeneously broadened spin spectra, especially at low bias field. As discussed above, owing to $d_\perp \gg d_\parallel$, the perpendicular component of the random electric field has the greatest effect, pushing the spin transition frequencies outward from the zero-field splitting. Figure~\ref{fig3}(a) illustrates the evolution in lineshape as an axial bias field is increased from 100 to 400~$\mu$T, showing only the $m_s = 0 \leftrightarrow m_s = +1$  transitions for simplicity. The spectrum for the $m_s = 0 \leftrightarrow m_s = -1$ transitions is approximately a mirror image about $\Delta = 2870$~MHz. Even for moderate bias field $\sim 1$~mT (see Fig.~\ref{fig3}(b)), a clear asymmetry persists, such that the spectrum is poorly fit by the commonly used symmetric Voigt function. 

To adequately fit asymmetric lineshapes, a better model is needed. %
To find a physically motivated function, we consider a large axial magnetic field, where the $m_s = 0 \leftrightarrow m_s=\pm1$ spin transition frequencies without hyperfine structure are given by 
\begin{eqnarray}
    f_\pm &=& \Delta \pm \sqrt{(\gamma_{NV} B_z)^2 + (d_\perp E_\perp)^2}\\
    &\approx& \Delta \pm \left(|\gamma_{NV} B_z| + \frac{(d_\perp E_\perp)^2}{2 |\gamma_{NV} B_z|}\right).\label{eq:approxf}
\end{eqnarray}
If we approximate
 $E_x$ and $E_y$ as Cauchy distributed and assume $B_z$ is also Cauchy-distributed with nonzero mean, a closed-form approximation for the probability distribution of spin transition frequencies can be found (see Appendix~\ref{app:ABCDE}):
 \begin{widetext}
     \begin{equation}
         P(z) \approx Im\left\{ \frac{\mathcal{B}_0}{\pi^2 (\mathcal{B}_0 \eta - \sigma^2)} \left(\frac{2\sigma \left[\pi - 2 \sin^{-1}\left(\frac{\sigma}{\sqrt{2\eta \mathcal{B}_0}}\right)\right]}{\sqrt{2 \mathcal{B}_0 \eta - \sigma^2}} - \pi\right)
        \right\},
    \end{equation}
    \end{widetext}
 where $z$ is the frequency $f_+ - \Delta$, $\eta = \mathcal{B}_0 - z + i\beta$, $\sigma$ is the HWHM of the Cauchy distribution of electric field components times $d_\perp$, and $\beta = 1/(2\pi T_2^*)$ and $\mathcal{B}_0 = \gamma_{NV}\langle B_z\rangle $ are the HWHM and mean, respectively, of the Cauchy distribution of axial magnetic fields expressed in frequency. The resulting Axial B-field Cauchy-Distributed E-field (``ABCDE'') model provides a much better fit to the simulated spectra (see Fig.~\ref{fig3}(b)), matching its asymmetry and allowing accurate estimation of the peak location. %

The second key feature introduced by internal electric fields is an outward shift of the lineshape peak, which can lead to systematic errors even if the peak location is accurately estimated. The nominal ($E = 0$) spin transition frequency in an axial field ($f_0 = \Delta + \gamma_{NV} B_z$, shown by dashed lines in Fig.~\ref{fig3}(a) and the blue  line in Fig.~\ref{fig3}(b)) consistently lies at lower frequency than the peak of the spectrum. As a result, if one were to naively use the peak of the spectrum $f_\text{peak}$ to estimate the magnetic field via $B_z = (1/\gamma_{NV})(f_\text{peak} - \Delta)$, the presence of RIEFs would induce a systematic error. Using the center of a fit with a Voigt function ($f_\text{Voigt}$) would produce an even larger systematic error because the Voigt function shifts further outward to attempt to match the asymmetric lineshape. 

The ABCDE model can partially mitigate the systematic error caused by RIEF-induced peak shifts. Beyond identifying the peak, the three parameters $\{\mathcal{B}_0, \beta, \sigma\}$ extracted by fitting to the ABCDE model have physical meaning; in particular, $\mathcal{B}_0$ provides an estimate for the Zeeman splitting in the presence of RIEFs and does not, in general, coincide with the maximum of the function. %
Figure~\ref{fig3}(b) illustrates the three frequencies $f_\text{peak}, f_\text{Voigt}, f_\text{ABCDE}$ with dashed lines, showing that $f_\text{ABCDE} = \mathcal{B}_0 +\Delta$ is considerably closer to the nominal spin transition frequency $f_0$ (solid blue line). %

From examination of %
Eq.~\ref{eq:approxf}, one might expect that the RIEF-induced systematic errors fall off %
inversely with $B_z$. Figure~\ref{fig3}(c) examines the magnetic field dependence of these systematic errors when the three different methods ($f_\text{peak}, f_\text{Voigt}, f_\text{ABCDE}$) are used to estimate the spin transition frequency $f_0$. While all three methods produce errors that fall off with magnetic field, the rate of decrease is far slower than expected, falling off approximately as $1/\sqrt{B_z}$. Qualitatively, as $B_z$ increases, it compresses the power-law tail in the spectral distribution, which shifts the peak of the distribution slightly outward. The impact of electric fields on magnetometry thus persists to larger bias fields than might naively be expected.

The systematic errors are significant, especially at low bias field or when $f_\text{peak}$ or $f_\text{Voigt}$ is employed. For typical bias fields of a few mT, and for dopant densities typical of commercially available samples, systematic errors above 1~$\mu$T are observed with $f_\text{peak}$ and $f_\text{Voigt}$, exceeding the systematic error coming from uncertainty in literature values for $g \approx 2.0028\pm 0.0003$ ~\cite{loubser_electron_1978}. These systematics also exceed the few-hundred-nT shifts associated with nonsecular hyperfine interactions in the few-mT-bias regime (see Appendix~\ref{app:hyperfine}). Notably, the ABCDE fit provides roughly an order of magnitude improvement over the Voigt fit. It is not perfect because the Monte Carlo simulation of internal electric fields is not precisely Cauchy-distributed, and because its approximations are only valid in the limit that $\gamma_{NV} B_z \gg d_\perp E_\perp$, which is not well satisfied at low bias. Nevertheless, for typical few-mT bias fields the error from the ABCDE fit is comparable to the current $g$-factor uncertainty.

\subsection{Concentration dependence}

\begin{figure}[htb]
\includegraphics[width=\linewidth]{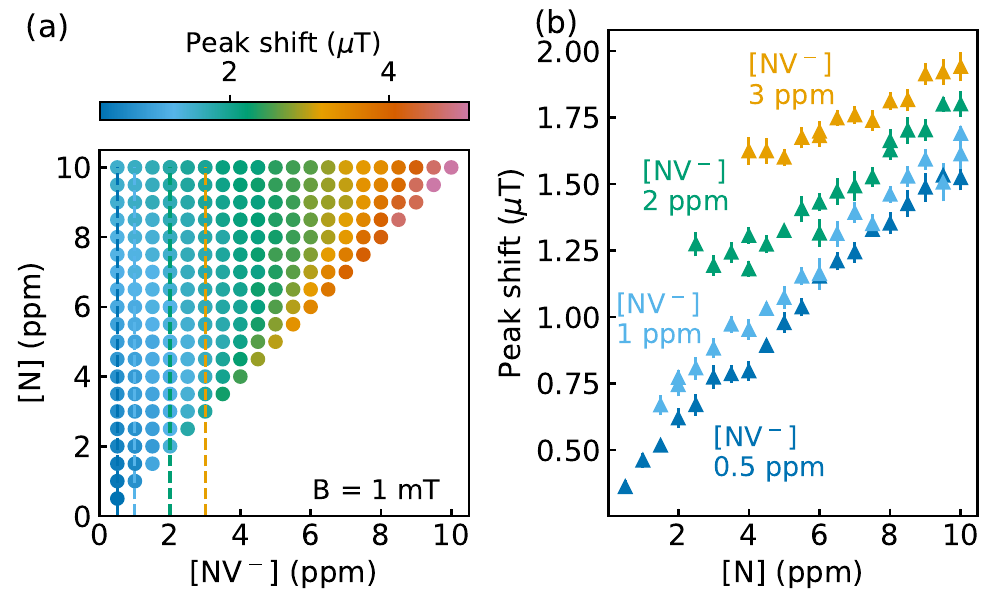}%
\caption{\label{fig4} (a) Variation in the peak shift $(f_\text{peak}-f_0)/\gamma_{NV}$ vs NV and N concentration in an axial magnetic field of 1 mT. Dashed lines show cuts plotted in (b). (b) Variation of peak shift with nitrogen concentration shows approximately linear behavior. In all panels, the spectrum for the $m_s = 0 \leftrightarrow m_s = +1$ transition is calculated from 100,000 electric field samples generated for specified concentrations and 100,000 magnetic field values generated from $T_2^* = 1.24~\mu$s. Spectra are generated using 501 bins over a 3 MHz range centered at $f_0$ and fit with the ABCDE model to extract the peak of the spectrum. Error bars are calculated with 50 bootstrap resamples (see Appendix~\ref{app:bootstrapping}).} %
\end{figure}

The magnitude of the RIEF-induced peak shifts varies dramatically with the concentration of dopants in the diamond. %
Figure~\ref{fig4} illustrates how systematic errors behave as a function of [NV$^-$] and [N]; we plot the shift in the peak of the spectrum from its nominal value, expressed in units of magnetic field, i.e., $(f_\text{peak}-f_0)/\gamma_{NV}$ for the $m_s = 0 \leftrightarrow m_s = +1$ transition in an axial magnetic field of $B_z = 1~$mT. 
These simulations assume a constant $T_2^* = 1.24~\mu$s, which would be appropriate if, for example, $T_2^*$ were limited by $^{13}$C nuclear spins or inhomogeneous strain rather than neutral nitrogen or other NVs. While variations in $T_2^*$ are likely to occur, we neglect them here for simplicity.
Our simulations exhibit a peak shift that increases approximately linearly with nitrogen concentration provided that [N] significantly exceeds [NV$^-$] (see Fig.~\ref{fig4}(b)). Deviations from this behavior for [NV$^-$] approaching [N] are an unphysical result of nearly every nitrogen becoming ionized in the simulation; in real samples the [NV$^-$]:[NV$^0$] ratio would likely decrease.  Overall, the simulations shown in Fig.~\ref{fig4} indicate that dilute samples with low [NV$^-$] and [N] would offer more accurate field estimation.

\subsection{Microwave broadening}

Thus far, we have considered only the bare distribution of spin transition frequencies. Perhaps surprisingly, additional broadening mechanisms can cause further shifts in the peak of the spectrum due to the underlying asymmetric distributions. 
If we begin with an inhomogeneous distribution of spin transition frequencies, say $s(f_0)$, and each spin transition frequency is broadened (e.g., due to microwave and/or optical excitation) by a homogeneous lineshape $h(f-f_0)$, the resulting signal is the sum of all the homogeneous lineshapes, i.e., the convolution $s * h$. Typical homogeneous lineshapes are symmetric (Lorentzian for continuous-wave microwave broadening), and convolution of an asymmetric $s(x)$ with a symmetric $h(x)$ will produce a spectrum that shifts to the long-tailed side of $s(x)$, i.e., in this case moving the peak further outward from the zero-field splitting.

\begin{figure}[htb]
\includegraphics[width=\linewidth]{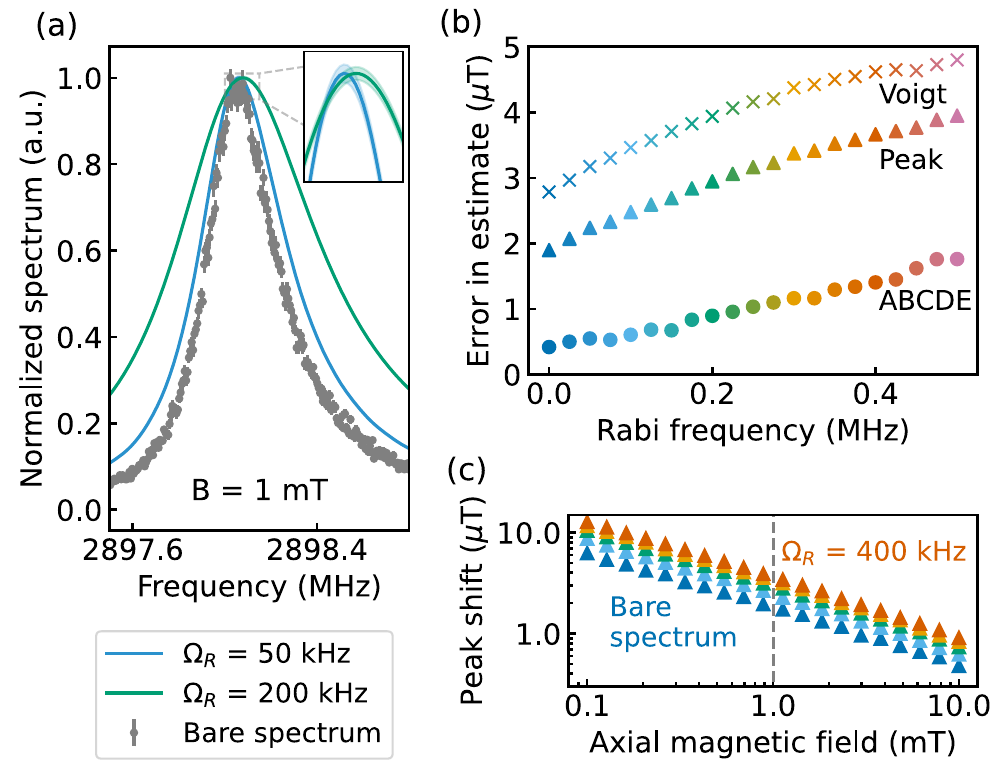}%
\caption{\label{fig5} (a) Microwave broadening, modeled as a Lorentzian with FWHM = 2$\Omega_R$, causes a shift in the peak position due to asymmetry in the underlying inhomogeneous distribution of transition frequencies. Example spectrum shown for 1 mT axial field. Inset shows peak shift between $\Omega_R = 50$ and 200 kHz (shaded areas are statistical uncertainty from the underlying Monte Carlo simulation). (b) Error in estimated field vs Rabi frequency at 1 mT axial field, as extracted via Voigt or ABCDE fit, or the peak in the ABCDE fit. (c) Shift in the peak position for different Rabi frequencies (color corresponds to colors in (b)) as a function of magnetic field. The dashed line indicates the field for which (b) is calculated.  In (b) and (c), error bars are smaller than the symbols.  In all parts, [NV$^-$] = 2.7 ppm, [N] = 8.5 ppm, and $T_2^*$ = 1.24~$\mu$s. %
}
\end{figure}

Figure~\ref{fig5}(a) illustrates the peak shift arising from convolution of a bare $m_s = 0 \leftrightarrow m_s = +1$ Monte Carlo spectrum with a Lorentzian corresponding to different MW Rabi frequencies. The peak of the broadened spectrum visibly shifts as the Rabi frequency is increased. To quantify the magnitude of these shifts, we fit the broadened spectrum with a Voigt or ABCDE model and extract $f_\text{Voigt}$, $f_\text{peak}$,  and $f_\text{ABCDE}$ as before and compare them to $f_0$.  The resulting error in the estimated field is shown as a function of Rabi frequency in Fig.~\ref{fig5}(b) for simulations with $B_z = 1$ mT. The errors vary by a similar factor across a range of magnetic fields (see Fig.~\ref{fig5}(c)), showing that Fig.~\ref{fig5}(b) is representative of typical behavior. The estimated magnetic field varies by 1--2 $\mu$T as the Rabi frequency increases from 0 to 500 kHz. This result illustrates that microwave power stability could become important for low-drift devices seeking $<10$~nT accuracy, where Rabi frequency stability at the few-kHz level could be required.

\section{Vector magnetometry}
Thus far, we have examined the simplest case of a single NV orientation in an axial magnetic field. Vector magnetometry, however, typically employs multiple orientations in the ensemble, which necessarily experience off-axis fields. We thus seek to understand how the RIEF-induced systematic errors observed in axial field sensing affect the vector-sensing case. 

In an ensemble, signals from all four NV orientations can be combined to measure the vector components of the magnetic field (within some permutation symmetries~\cite{mongelos-martinez_unconditional_2026}). A standard approach extracts each orientation's spin transition frequencies from an ODMR spectrum, and compares them with the eigenvalues of the four Hamiltonians (Eq.~\ref{eq:H}) describing the different orientations ~\cite{schloss_simultaneous_2018, garsi_three-dimensional_2024, lonard_limits_2026}. If the susceptibilities $d_\parallel, d_\perp$, the gyromagnetic ratios  $\gamma_{NV}, \gamma_N$, and hyperfine parameters $P, A_\parallel, A_\perp$ are known, and the environment of the ensemble can be described by a single homogeneous electric and magnetic field, then just 7 real parameters remain undetermined: $\Delta, B_x^C, B_y^C, B_z^C, E_x^C, E_y^C, E_z^C$. Here, the superscript $C$ indicates that vector components are defined in the conventional diamond crystallographic coordinate system. Each Hamiltonian is evaluated after transforming these components into the corresponding NV's frame.
Alternatively, some approaches assume $E_x, E_y$ are the same for all orientations and subsume $E_z$ into $\Delta$, reducing the number of free parameters to 6~\cite{garsi_three-dimensional_2024}. The 8 electron spin transition frequencies overdetermine the 6 or 7 unknowns, so $B_x^C, B_y^C, B_z^C$ follow from least-squares minimization of
\begin{equation}
\label{eq:chi2}
\chi^2(\mathbf{B}^C,\mathbf{E}^C,\Delta)=\sum_{i=1}^{4}\sum_{m_s=\pm 1}\left[\nu_{m_s}(\mathbf{M}_i,\mathbf{B}^C,\mathbf{E}^C,\Delta)-f_{i,m_s}\right]^2
\end{equation}
where $\nu_{m_s}$ are the eigenvalues of the Hamiltonian, $\mathbf{M}_i$ is the transformation matrix for each orientation, and $f_{i,m_s}$ is the spin transition frequency extracted from fitting the spectrum.

Such an analysis neglects random internal electric fields. First, spin transition frequencies are typically estimated by fitting a Lorentzian or Voigt profile. While the underlying asymmetry of the inhomogeneously broadened spin spectrum may be masked by measurement noise or significant microwave or optical broadening, a symmetric fit function can nevertheless introduce systematic error (as seen in the MW-broadened fits in Section~\ref{sec:single}). Second, including a homogeneous electric field or an identical internal field for each orientation may not adequately mimic the effects of inhomogeneous electric fields, potentially further distorting the estimate of $\mathbf{B}^C$.

\begin{figure}[htb]
\includegraphics[width=\linewidth]{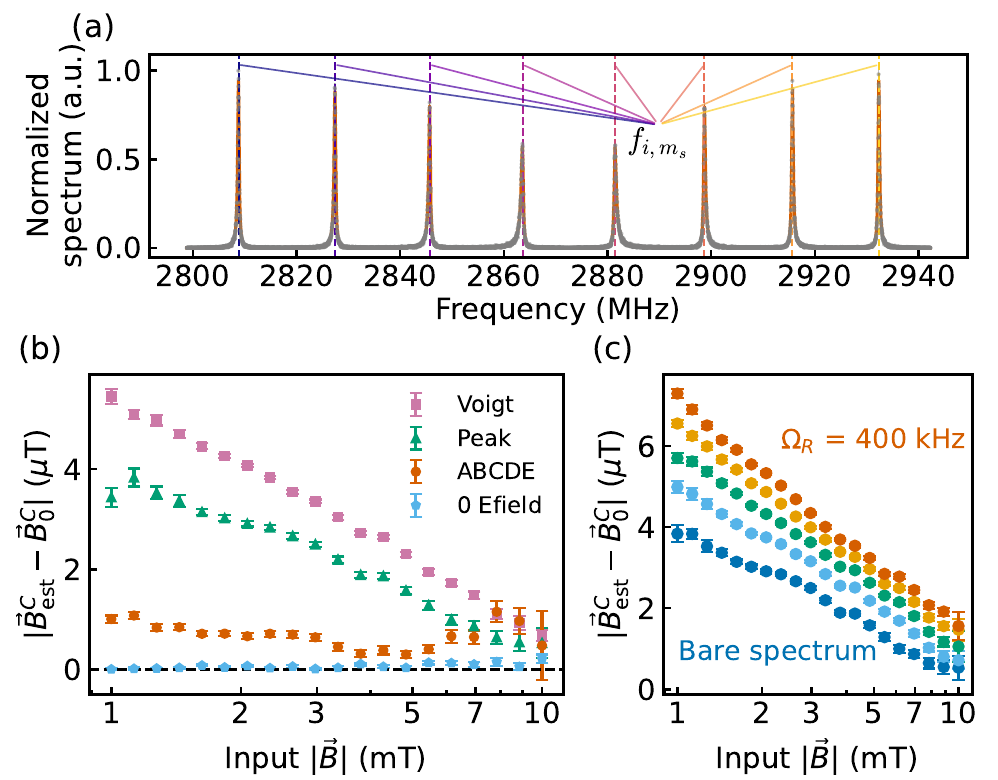}
\caption{\label{fig6} (a) Example ODMR spectrum for the $\langle 4~2~1 \rangle$ bias field, showing all four NV orientations, with ABCDE fit overlaid. The 8 extracted frequencies are compared via a cost function to the eigenvalues of the 4 Hamiltonians. (b) Error in estimated vector magnetic field $|\mathbf{B}^C_\text{est} - \mathbf{B}^C_0|$ vs input field magnitude $|B|$, extracted via Voigt and ABCDE fits with 7 free parameters. The ``control" case, in which the simulated data contain no internal electric fields, is shown and recovers $\mathbf{B}^C_0$ to within fit uncertainty. (c) Change in the accuracy for different Rabi frequencies as a function of magnetic field, found via ABCDE peak values. In all parts, [NV$^-$] = 2.7 ppm, [N] = 8.5 ppm, $T_2^* = 1.24~\mu$s, and $\Delta = 2870$ MHz. %
}
\end{figure}

To understand the magnitude of systematic errors introduced by this type of analysis, we apply it to simulated data sets that include internal electric fields and compare the estimated magnetic field $\mathbf{B}^C_\text{est}$ to the ground truth $\mathbf{B}^C_0$ used in our simulation. As above, we consider an ensemble with [NV$^-$] = 2.7 ppm, [N] = 8.5 ppm, $T_2^* = 1.24~\mu$s, and $\Delta = 2870~$MHz, and neglect hyperfine structure. We choose an external magnetic field along the $\langle4~2~1\rangle$ direction in diamond, as this choice ensures equal spacing between ODMR features from each orientation and thus is commonly employed for biasing NV vector magnetometers. Figure~\ref{fig6}(a) shows a representative simulated spectrum for this geometry at $|\mathbf{B}| = 2.5$~mT. The eight resolved resonances correspond to the $m_s = \pm1$ transitions of the four orientations; dashed lines mark the extracted frequencies $f_{i,m_s}$, which are compared against the eigenvalues of all four Hamiltonians in the $\chi^2$ sum of Eq.~\ref{eq:chi2}. We apply the analysis described above using a Voigt fit and assuming seven free parameters: $\Delta, \mathbf{B}^C, \mathbf{E}^C$. For comparison, we apply the same analysis using an ABCDE fit to estimate the spin transition frequency using both $\mathcal{B}_0 + \Delta$ and the peak of the ABCDE fit. While the assumptions underlying the ABCDE model do not strictly apply in off-axis fields, it may nevertheless provide an improvement over a Voigt fit.

Figure~\ref{fig6}(b) shows the magnitude of the systematic errors in the reconstructed vector magnetic field for three methods of extracting the spin transition frequencies, along with a control case. To ensure the reconstruction works as intended, we first apply it to simulated spectra containing no internal electric fields, and find that the control case recovers $\mathbf{B}^C_0$ to within the fit uncertainty. With internal electric fields, the peak shifts revealed in the single-axis case propagate through our analysis, causing systematic errors as large as $\sim5~\mu$T in the reconstructed vector field, particularly at low field ($|\mathbf{B}| \approx 1$~mT). In this low-field regime, both ABCDE-based frequency estimates improve on the Voigt profile. As expected from the single-axis case, accuracy improves with increasing field strength as the Zeeman splitting comes to dominate the RIEF-induced peak shifts. Unlike the single-axis case, however, the Voigt estimate improves more rapidly than the ABCDE estimates, and all three converge at $\sim8$~mT. We attribute this to the axial-field assumption underlying the ABCDE model, which cannot hold for every orientation when all four are probed simultaneously. At low field this is outweighed by the model's correct treatment of the inhomogeneous electric field distribution, but as a large magnetic field suppresses the transverse electric field contribution, the approximation error dominates and the ABCDE advantage diminishes. 

MW broadening worsens the reconstruction accuracy similarly to the single-axis case (Fig.~\ref{fig6}(c)), with the error increasing uniformly with $\Omega_R$ at all field strengths while the field dependence is preserved. Notably, combining signals from four orientations does not mitigate the errors seen in the single-axis case. Indeed, for a given $|\mathbf{B}|$, we find larger errors than in the single-axis case since some orientations inevitably see a small axial field, making their RIEF-induced shifts comparatively large. As in the single-axis case, the resulting $\mu$T-scale RIEF-induced errors exceed those from uncertainty in $g$, and could be experimentally detectable in a known and stable reference field.

\section{Conclusion}

Using Monte Carlo simulations, we have examined the role that internal electric fields can play in NV magnetometry. First, RIEFs can quantitatively explain the asymmetric spin-resonance lineshapes observed in finite-bias ODMR, providing evidence for their importance when analyzing low-noise data. Second, RIEFs cause a shift in the peak of the spin resonance lineshape that falls off much more slowly than $1/B_z$, representing a source of inaccuracy at the $0.1\%$ level for typical defect densities and few-mT fields.   

Larger bias fields could suppress RIEF-induced shifts, offering one mitigation strategy. However, drift in the bias field itself would have a proportionally larger impact on accuracy. Recent work has proposed bias-field-free experiments to eliminate bias-drift-induced inaccuracy~\cite{childress_bias-field-free_2025}, an approach that would render the magnetometer highly susceptible to RIEF-induced systematic errors. Reduced defect densities offer a more appealing solution to decreasing RIEF-induced shifts; in sufficiently clean materials, reduced NV and N concentrations can also increase $T_2^*$ and improve sensitivity~\cite{barry_sensitivity_2020}. Finally, the ABCDE model offers up to an order-of-magnitude improvement over a Voigt lineshape fit in estimating magnetic fields from asymmetric lineshape data. 

We have shown that RIEFs can form a substantial portion of the systematic error budget in NV ensemble magnetometry, but to reach a competitive absolute accuracy many other factors will need to be considered. A detailed understanding of the solid-state environment will be required, including not just RIEFs and nonsecular $^{14}$N hyperfine interactions (see Appendix~\ref{app:hyperfine}), but possibly also inhomogeneous strain~\cite{marshall_high-precision_2022} and mosaicity, interactions with paramagnetic dopants~\cite{simanovskaia_sidebands_2013}, $^{13}$C, phonons, and more. Moreover, our electric field model itself is relatively crude: it approximates screening effects through a dielectric constant~\cite{alaerts_first-principles_2024}, assumes homogeneous defect distributions~\cite{shao_statistical_2026}, and neglects optically induced dynamics in the charge distribution~\cite{aslam_photo-induced_2013, jayakumar_optical_2016}, all of which may become important as accuracy improves. Crucially, these models will need to be compared to experiments with precisely controlled magnetic field environments using new data-analysis approaches, as brute-force least-squares fitting of complex models may be too slow for real-world applications.

While absolute accuracy remains a significant challenge, a better understanding of the physical mechanisms underlying NV response to magnetic fields could also potentially improve empirical calibrations. For example, we observe RIEF-induced dependence of spin transition frequencies on Rabi frequency, which could spur inclusion of MW power in training datasets. Similarly, an explanation for asymmetric lineshapes that complicate calibrations could provide a route to mitigating them by reducing sample dopant density. Conversely, dependencies revealed in empirical calibrations could provide clues to missing physics in current first-principles NV magnetometer analysis. Indeed, pushing the envelope on accuracy opens an opportunity to better connect our understanding of the fundamental physics of the NV center to its applications in magnetometry.

\begin{acknowledgments}
 The authors acknowledge use of LLMs (Claude Sonnet 4.6-5.5) to assist in code creation by generating code snippets, for example for layouts and formatting of figures, vectorization of functions, syntax of fitting packages and similar low-level functionalities, which were then human-edited and incorporated into the codebase. In addition, the code to illustrate the lattice points in Fig.~\ref{fig1}(a) was AI-generated and human-edited. Code was written in a workspace without direct AI access; the code architecture as well as the underlying mathematical and physical theory was human designed. The manuscript was entirely human-written and edited. 

 JF thanks Kenny Moc for helpful discussions on cost function optimization. This work was carried out as part of grants QSP-051 and QSP-125 from the National Research Council Quantum Sensing Challenge Program; DG was supported by the McGill SURA (Science Undergraduate Research Award) program; LC also acknowledges support from NSERC Discovery RGPIN-2026-04512 and DND supplement DGDND-2026-04512. JF and LC benefit from membership in Quebec strategic groups RQMP (https://doi.org/10.69777/309032) and INTRIQ.
\end{acknowledgments}

\appendix

\section{Monte Carlo simulation of ODMR spectra \label{app:MC}}

\subsection{Lattice construction}
\label{app:MC_lattice}

Our simulation is based on placement of charged dopants within the diamond lattice. First, substitutional nitrogen is homogeneously and isotropically distributed on diamond lattice sites in a cubic volume. NV$^-$ centers are then placed sequentially at random into the lattice, re-sampling any placement within 1.2 nm of  N~\cite{manson_nvn_2018}. For each NV$^-$, we assume that the additional electron comes from the closest neutral nitrogen, which is thereby converted to N+. We also assume that both the spin density and excess charge of the NV$^-$ are localized on the vacancy ~\cite{he_paramagnetic_1993}, and that screening effects are adequately described by relative permittivity ~\cite{alaerts_first-principles_2024}. Since we are only interested in the electric field generated by charged dopants, we neglect NV0.

For most simulations we use a cubic volume of $m^3$ conventional unit cells with m = 400; for data shown in Fig.~\ref{fig4} with [N] $\geq$ 3~ppm we use m = 200 to speed computation times. %
To ensure that we were using a sufficiently large volume, we compared spectra as a function of $m$, looking for the point at which further reductions in $m$ led to significant changes in the spectrum (i.e. larger than the uncertainty in the MC estimate). The uncertainty in the spin spectrum was estimated by subdividing 100,000 instances into 50 subsets of 2,000 instances, whose transition frequencies were histogrammed to create 50 spectra. The final spectrum is the mean of these spectra, and each bin's uncertainty is given by $\sigma/\sqrt{50}$. We then calculated the reduced $\chi^2$ representing the difference between the test ($m_{test} = m$) and target ($m_{target} = 400$) spectrum as a function of $m$ and found no statistically significant difference between the two spectra for $m>100$ for all concentrations considered.

The randomized field calculation was implemented in Python with partial vectorization, resulting in minutes to hours run times for simulation of 100,000 field instances, depending on dopant concentrations (higher density requires more time); future versions incorporating just-in-time compilation and parallelization should permit considerable speedup. 

\subsection{Spectrum calculation \label{app:MC_unweighted}}

Spectra are computed from $N$ Monte Carlo electric-field instances ($N = 100{,}000$ for the fit in Fig.~2 and for Fig.~4; $N = 200{,}000$ for Figs.~1, 3, 5, and 6). For each instance, we also draw independent random offsets for $B_x, B_y$ and $B_z$ from a Cauchy distribution with HWHM given by $1/(2\pi T_2^*\gamma_{NV})$. These values, together with an applied DC magnetic field, are used as inputs to the $3\times3$ spin Hamiltonian (the DC overall electric field is set to zero for simplicity), producing two transition frequencies per instance, per NV orientation, and per nuclear spin projection $m_I$ (included in the secular approximation). Transition frequencies are calculated using an extension of the analytical formulas provided in~\cite{lonard_limits_2026}, where we treat $E_x$ and $E_y$ separately since the transverse angle of the electric field matters when the transverse magnetic field is nonzero. MW broadening is approximated by summing Lorentzians $\Omega_R^2/[2(\delta^2 + \Omega_R^2)]$, with Rabi frequency $\Omega_R$ and detuning $\delta$, centered on each histogram bin and weighted by its counts, which closely approximates a convolution. The edges of the spectrum (within $3\Omega_R$ of the edge) were excluded from the fits to mitigate edge effects. 
The uncertainty in MW-broadened spectra (shown in Fig. 5(a) inset) is found by standard Gaussian uncertainty propagation from the 50-subsample standard error in the underlying inhomogeneous spectrum. 

Using vectorized computations in Python, we perform secular model simulations (similar to those shown in Fig.~2) from a pre-generated distribution of 100,000 electric fields in around one second for spectrum generation and a few seconds for MW broadening on a standard laptop (MacBook Pro, 2~GHz quad-core Intel Core i5). Because of the long computational time for the MC electric-field generation relative to spectrum generation, the same set of $N$ electric-field instances is reused to calculate spectra under different configurations (field, MW broadening). As a result, error bars in figures may exceed point-to-point fluctuations but nevertheless represent the true statistical uncertainty.

\section{Fitting the Monte Carlo simulation to experimental data}

\begin{figure}[htb]
\includegraphics[width=\linewidth]{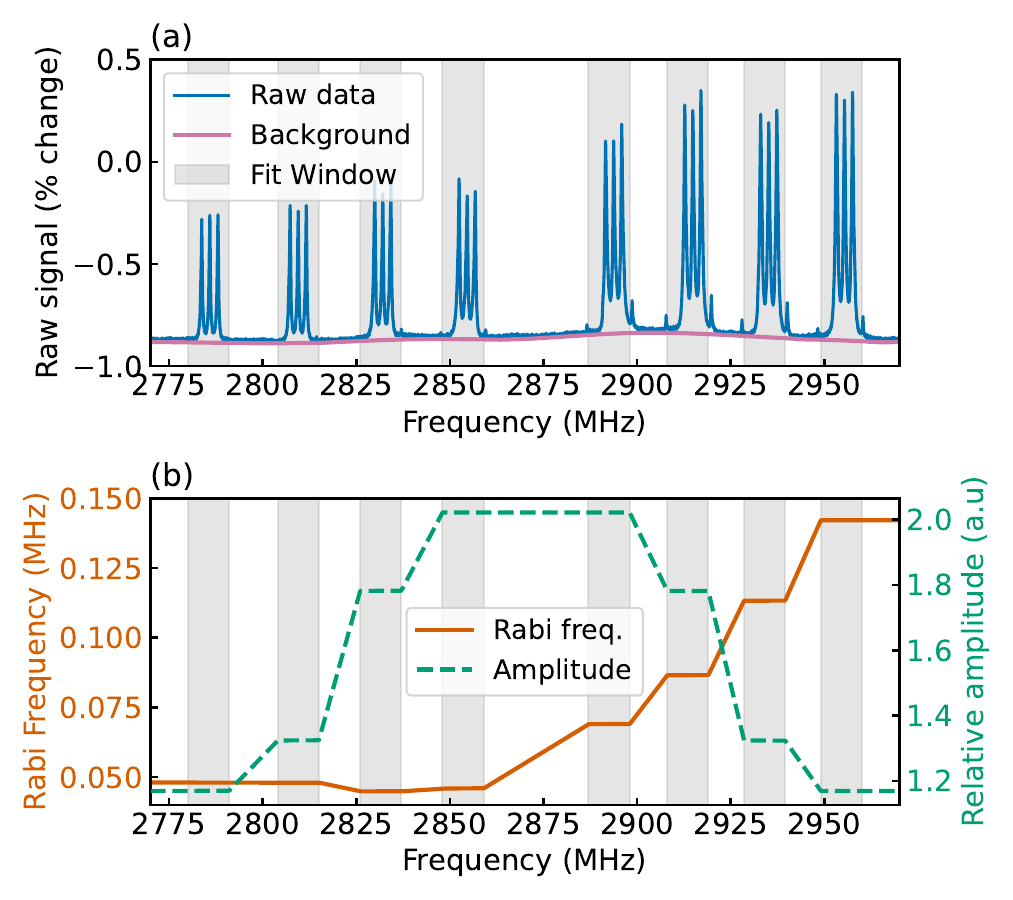}
\caption{\label{fig:fig7} (a) Data preparation. The raw voltage signal from the photo-detection circuit exhibited a background drift (purple), which was subtracted from the data. Gray shaded regions indicate the portions of the data that were fit to the model. (b) The Rabi frequency (left) and relative amplitude (right) extracted for each triplet from the fit to the secular model.}
\end{figure}

We fit the secular spectrum model to experimental data as follows. We first remove a slow drift in the photodiode signal from the data using the Baseline.snip method from the \texttt{pybaselines} package~\cite{erb_pybaselines_2025} (see Fig.~\ref{fig:fig7}(a)), and normalize the data to lie between 0 and 1. Since we are fitting to a secular model that does not include weakly allowed nuclear-spin-nonconserving transitions, we crop the data around each of the triplets, excluding the largest satellite peaks (see gray regions in Fig.~\ref{fig:fig7}). With [NV$-$] and [N] concentrations fixed to the nominal levels for the diamond sample used in the magnetometer, we optimize over other parameters including the zero-field splitting ($\Delta$), the magnetic field (${B_x, B_y, B_z}$), magnetic broadening ($T_2^*$), and an offset. Furthermore, to account for variations in MW power with frequency induced by the response of the MW resonator, we consider 8 Rabi frequencies, one for each triplet. The varying Rabi frequency affects both the broadening and the contrast. We also allow the 4 amplitudes of each orientation's resonance to vary to account for an unknown optical polarization. In total, there are 18 real-valued fit parameters.

We fit using the Covariance Matrix Adaptation Evolution Strategy (CMA-ES) optimization algorithm as implemented in the \texttt{cma} Python package~\cite{hansen_cma-espycma_2019}; this derivative-free algorithm is well-suited to noisy models with a large number of parameters. Nevertheless, it was necessary to fine-tune the initial conditions by hand to allow fitting within a day timescale. Fit parameters were rescaled to units of MHz, $\mu$s or gauss (so that their values were of order unity) and an initial step size of $\sigma_0  = 0.001$ was employed. The parameters used in Fig.~\ref{fig2} were selected from the CMA-ES run that gave the best agreement with the data: $T_2^* = 1.24477~\mu$s, $\Delta = 2869.1983$ MHz, and $\mathbf{B} = \{3.22173, 1.34076, 0.67760\}$~mT. Note that it is not expected that this magnetic field value will be highly accurate because the model does not include nonsecular hyperfine interactions (see Appendix~\ref{app:hyperfine}). Because the goal of this fit was to demonstrate that the model reproduced asymmetric lineshapes observed in the data (rather than to extract precise parameters), evolution was stopped after 5000 iterations.

\label{app:fit}

\section{Deterministic models}
\label{app:ABCDE}
\subsection{The nearest-N$^+$ model}

Figure~\ref{fig1}(c) includes an N$^+$-only probability distribution for $E_\perp$ that is deterministically (though not analytically) generated. Ref.~\cite{yu_optically_2024} developed a model for the electric field produced solely by ionization of the closest nitrogen, assuming a homogeneous and isotropic distribution of dopants. We reproduce a portion of their derivation here and explain the adaptation we make. First, we find the probability $P(r)\,dr$ that the nitrogen closest to the NV center of interest lies between $r$ and $r + dr$. This is the product of the probability $A_0$ that there is no nitrogen closer than $r$ and the probability $B_1$ that there is at least one nitrogen between $r$ and $r + dr$. Assuming a Poisson distribution for the number of nitrogen dopants in a given volume, the probability $P_k(V)$ of having $k$ nitrogen dopants in a volume $V$ is $\lambda^k e^{-\lambda}/k!$, where $\lambda = \rho V$ and $\rho$ is the density of nitrogen. Thus $A_0 = P_0(4\pi r^3/3) = e^{-4\pi r^3 \rho/3}$ and $B_1 = 1-P_0(4\pi r^2 dr) = 1 - e^{-4\pi r^2 \rho dr} \rightarrow 4\pi r^2 \rho dr$ in the limit $dr \rightarrow 0$. Consequently,
\begin{equation}
    P(r) = 4\pi r^2 \rho e^{-4\pi r^3\rho/3}.
\end{equation}
The magnitude of the electric field, $|E| = e/(4\pi\epsilon_0 \epsilon_r r^2) = \alpha^2/r^2$, has a probability distribution $P_{E}(E) = P(r)/|dE/dr| = P(r) r/(2 E)$ or
\begin{equation}
    P_{E}(E)= 2\pi \alpha^3 \rho E^{-5/2} e^{-4\pi \alpha^3 \rho E^{-3/2}/3}.
\end{equation}
While Ref.~\cite{yu_optically_2024} presumes that the distribution of the perpendicular component of the electric field is similar to that of $|E|$ (i.e., $E_\perp\propto E$), we find that this assumption produces a qualitatively different distribution than explicitly accounting for the angle between the field and the NV-axis. Instead, we set $E_\perp = E\sin{\theta}$, where $\theta$ is the random angle between the electric field and the NV axis, and note that $x = \sin{\theta}$ has the probability distribution $P_x(x) = x/\sqrt{1-x^2}$ when the field direction is isotropically distributed over $4\pi$ solid angle. We can then find the probability distribution for $E_\perp$ by integrating over x:
\begin{equation}
     P_{E_\perp}(E_\perp) = \int_0^1 P_E\left(\frac{E_\perp}{x}\right)\frac{1}{\sqrt{1-x^2}}\,dx.
\end{equation}
In Fig.~\ref{fig1} we show the resulting expression evaluated for the nominal nitrogen density of 8.5~ppm. Unsurprisingly, it underestimates the electric field because it does not include contributions from other NV$^-$ or N$^+$ present in the lattice.

\subsection{The ABCDE model}

In a large axial magnetic field, neglecting hyperfine interactions, the spin transition frequencies between the $m_s = 0$ and $m_s = \pm1$ states for Eq.~\ref{eq:H} are given by
\begin{eqnarray}
    f_\pm &\approx& \Delta \pm \left(\gamma_{NV} |B_z| + \frac{(d_\perp E_\perp)^2}{2 \gamma_{NV} |B_z|}\right),
\end{eqnarray}
where the approximation is quantitatively valid for $\gamma_{NV}B_z \gg d_\perp E_\perp$. Recasting our variables to frequency units with $\mathcal{E}_\perp = d_\perp E_\perp$ and $\mathcal{B} = \gamma_{NV} B_z$, and treating both $\mathcal{E}_\perp$ and $\mathcal{B}$ as random variables, we approximate
\begin{equation} \label{eq:fABCDE}
    f_+  - \Delta \approx |\mathcal{B}| + \frac{\mathcal{E}_\perp^2}{2\langle|\mathcal{B}|\rangle},
\end{equation}
where we have replaced $\mathcal{B}$ by its mean value in the denominator; this replacement would be appropriate for $\langle \mathcal{B}\rangle$ much larger than the variations in $\mathcal{B}$, so it is a questionable approximation for a Cauchy-distributed random variable with infinite variance, yet it seems to provide reasonably good performance. 
Next, we define probability distributions for the two terms: $|\mathcal{B}|$ and $\mathcal{E}_\perp^2/2\langle\mathcal{B}\rangle$. We expect $|\mathcal{B}|$ to be approximately Cauchy distributed with mean $\mathcal{B}_0>0$ and HWHM $\beta\ll\mathcal{B}_0$: 
\begin{equation}
\label{eq:B}P_{|\mathcal{B}|}(x) = \frac{\beta}{\pi((x-\mathcal{B}_0)^2 + \beta^2) }.
\end{equation}
Here we have approximated the distribution of $|\mathcal{B}|$ with that for $\mathcal{B}$, which is valid in the limit $\mathcal{B}_0\gg\beta$. For the second term, we treat $\mathcal{E}_\perp^2 = \mathcal{E}_x^2 + \mathcal{E}_y^2$ as a random variable with $\mathcal{E}_x$ and $\mathcal{E}_y$ independent and Cauchy distributed, $P_{\mathcal{E}_i}(x) = \sigma/\pi(x^2 + \sigma^2)$. The probability distribution for $\mathcal{E}_\perp^2/2\langle\mathcal{B}\rangle$ is then 
\begin{equation}
\label{eq:Eperp2}
P_{\mathcal{E}_\perp^2/2\mathcal{B}_0} (x)= \frac{2 \sigma \mathcal{B}_0}{\pi\sqrt{\sigma^2 + 2 \mathcal{B}_0 x}(\sigma^2 + \mathcal{B}_0 x)}
\end{equation}
with domain $x> 0$.
The desired quantity (Eq.~\ref{eq:fABCDE}) is thus given by a convolution of Eq.~\ref{eq:Eperp2} with Eq.~\ref{eq:B}: 
\begin{equation}
    P_{f_+ - \Delta}(z) = \int_0^\infty P_\mathcal{|B|}(z-x)P_{\mathcal{E}_\perp^2/2\mathcal{B}_0} (x) ~dx,
\end{equation}
with the integration limits set by the domains of Eqs.~\ref{eq:B} and~\ref{eq:Eperp2}.

While unsightly, it is possible to find an analytic solution to the integral %
in explicitly real form; the most compact representation we found for $P_{f_+ - \Delta}(z)$ is
\begin{equation}
    \operatorname{Im}\left\{ \frac{\mathcal{B}_0}{\pi^2 (\mathcal{B}_0 \eta - \sigma^2)} \left(\frac{2\sigma \left[\pi - 2 \sin^{-1}\left(\frac{\sigma}{\sqrt{2\eta \mathcal{B}_0}}\right)\right]}{\sqrt{2 \mathcal{B}_0 \eta - \sigma^2}} - \pi\right)
    \right\}
\end{equation}
\noindent where $\eta = \mathcal{B}_0 - z + i\beta$.
This expression approximates the probability distribution of $f_+ - \Delta$, where $B>0$ and $z>0$. Because $f_-  - \Delta = -(f_+ - \Delta)$, the same expression can be employed for $f_-$ with appropriate signs.

\section{Fitting Voigt and ABCDE models to simulated data}\label{app:bootstrapping}
Fitting of simulated spectra to Voigt and ABCDE profiles was performed using least-squares minimization with the LMFit package. The ABCDE model is only well-defined for positive frequencies and positive $\mathcal{B}_0$ or negative frequencies and negative $\mathcal{B}_0$. Consequently, for ABCDE fitting, we subtracted $\Delta = 2870$~MHz from the simulated spectrum frequencies, and fit the positive and negative frequency halves separately. Reasonable estimates for the uncertainty in fit parameters (roughly in agreement with bootstrapping) could be found by weighted fitting to spectra, i.e. including the uncertainty in the spectrum. However, because the spectrum is most uncertain where the transition probabilities are high, weighted fits produced greater systematic error relative to the expected transition frequency as they prioritized matching the spectrum values away from the peaks. Because the fit functions are not perfect models for the data, the resulting fit did not match as well in the regions of the spectrum that carry the most information (the peaks). Consequently, all fits to spectra were done without weighting by the uncertainty in the simulated spectrum. Because of imperfect models, fit outcomes can also be influenced by the range of frequencies included in the fit. For Figs.~\ref{fig3} and \ref{fig4}, fits are performed over a frequency range of 3 MHz centered on the expected transition frequency $f_0$, which is sufficient to encompass the full peak. For Fig.~\ref{fig5}, we use a frequency range of 10 MHz $ -$ $ 6 \Omega_R$ centered on $f_0$, providing sufficient bandwidth to fit Rabi-broadened lines. The reduction in width by $3\Omega_R$ on each edge largely eliminates edge effects from the convolution.

To estimate uncertainties in $f_\text{peak}$, $f_\text{ABCDE}$ and $f_\text{Voigt}$, we used standard non-parametric bootstrapping~\cite{efron_introduction_1994}. The general process is as follows: 1.) Perform a fit with $x$ instances of the simulated electric field and extract a point estimate of the peak or, in the case of Fig.~\ref{fig6}, the reconstructed magnetic field. 2.) Resample the $x$ instances of the simulated electric fields (with replacement) and recalculate the peak position or magnetic field N times. 3.) The uncertainty in the estimate is found from the standard deviation of the recalculated values. Note that each resample also redraws the random magnetic-field offsets, so the quoted uncertainty reflects both E-field and magnetic-environment sampling. In Figs.~\ref{fig3}, \ref{fig5}, and \ref{fig6} we use N = 50 simulations each with $x = $200,000 instances of the simulated electric field. For Fig.~\ref{fig4}, N = 50 simulations are used, each with 100,000 field values.

\section{Systematic shifts from hyperfine interactions} 
\label{app:hyperfine}

To isolate the impact of internal electric fields, we have largely neglected the hyperfine interactions with the host nitrogen nucleus (typically $^{14}$N).  Here we discuss the effects of hyperfine interactions in the absence of internal electric fields. Hyperfine interactions are commonly treated within the secular approximation, which neglects the off-axis hyperfine terms ($A_\perp$); the nuclear Zeeman term is also often neglected. With $A_\perp = 0$ and $\gamma_N = 0$, the $m_I = 0$ transition frequencies are the same as the pure electronic spin transition frequencies. Thus we can map the hyperfine-free transitions modeled in the main text onto the $m_I = 0$ component of a secular treatment.

Neglecting nonsecular hyperfine interactions introduces systematic error in magnetometry. First, the resonance frequencies for nuclear-spin-conserving transitions are shifted slightly by the nuclear Zeeman and significantly by the off-axis hyperfine terms. Second, $A_\perp$ hybridizes the spin states such that nominally forbidden transitions become weakly allowed, producing satellite peaks that can distort secular fits to the spectrum. 

\begin{figure}[htb]
\includegraphics[width=\linewidth]{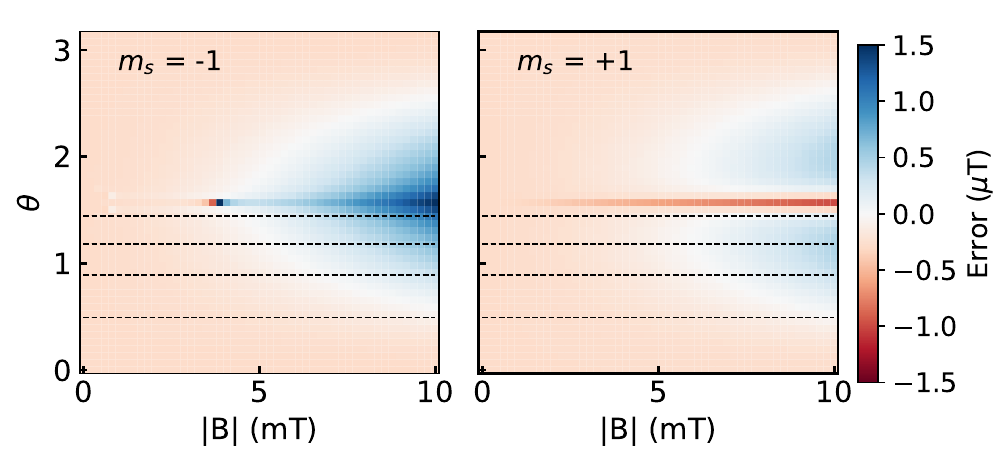}
\caption{\label{fig8} Systematic errors associated with neglecting hyperfine interactions for $m_s = 0 \leftrightarrow m_s = -1$ (left) and $m_s = 0 \leftrightarrow m_s = +1$ (right) as a function of external field magnitude $|B|$ and polar angle $\theta$ relative to the NV symmetry axis. Eigenvalues of Eq.~\ref{eq:H} with $\mathbf{E} = 0$ are calculated  for the $m_I = 0$ eigenstates with and without setting $A_\perp = 0$ and $\gamma_N = 0$, and the difference in transition frequencies is converted to $\mu$T via $\gamma_{NV}$. Dashed lines indicate polar angles for the four NV orientations in a $\langle 4~2~1\rangle$ magnetic field.}
\end{figure}

To quantify the size of the systematic errors introduced by the first mechanism, we examine the shift in the $m_I = 0$ peaks caused by nonzero $A_\perp$ and $\gamma_N$. The $m_I = 0$ peaks do not have any satellite peaks nearby and can thus be more reliably extracted than the $m_I = \pm 1$ peaks. Figure~\ref{fig8} shows the nonsecular hyperfine shift in $m_I = 0$ transition frequency (converted to magnetic field via $\gamma_{NV}$) for different magnetic field magnitudes and angles relative to the NV symmetry axis. These shifts are computed exactly from the eigenvalues of the Hamiltonian with and without the nonsecular hyperfine terms. Notably, the shift remains around a few hundred nT except for strong off-axis fields. Even for vector magnetometry, where typical bias fields are not aligned with the NV axis, truly perpendicular fields are not employed (see dashed lines in Fig.~\ref{fig8}). Thus for the field range of interest, up to a few mT bias field, systematic errors associated with hyperfine interactions are smaller than what we observe for electric-field-induced peak shifts in dense ensembles.

\bibliography{references}%

\end{document}